\documentclass[journal=nalefd,manuscript=article]{achemso}
\usepackage[version=3]{mhchem} 
\usepackage{float}
\usepackage{xcolor}
\usepackage{caption}
\usepackage{amsmath}
\usepackage{tabularx}
\usepackage{graphicx}
\usepackage{siunitx}
\usepackage{hyperref}
\usepackage{tikz}
\newcommand*\circled[1]{\tikz[baseline=(char.base)]{
            \node[shape=circle,draw,inner sep=2pt] (char) {#1};}}

\author{Atanu Ghosh}
\affiliation{Department of Physics, IIT Madras, Chennai, 600036, India}
\alsoaffiliation{Center For Quantum Information, Communication And Computing (CQuICC), Indian Institute of Technology Madras, Chennai, India 600036}
\altaffiliation{These authors contributed equally}

\author{Krishna Kumari Swain}
\affiliation{Department of Physics, IIT Madras, Chennai, 600036, India}
\alsoaffiliation{Department of Chemistry, Rajalakshmi Engineering College, Chennai, 602105, India}
\altaffiliation{These authors contributed equally}

\author{Agniva Das}
\affiliation{Department of Physics, IIT Madras, Chennai, 600036, India}

\author{Mrutyunjaya Rath}
\affiliation{Department of Physics, IIT Madras, Chennai, 600036, India}

\author{Snigdhadev Chakraborty}
\affiliation{Department of Physics, IIT Madras, Chennai, 600036, India}

\author{Bipeen Kumar}
\affiliation{Department of Physics, IIT Madras, Chennai, 600036, India}

\author{Yamini Selvam}
\affiliation{Department of Physics, IIT Madras, Chennai, 600036, India}

\author{Siddharth Dhomkar}
\affiliation{Department of Physics, IIT Madras, Chennai, 600036, India}
\alsoaffiliation{Center For Quantum Information, Communication And Computing (CQuICC), Indian Institute of Technology Madras, Chennai, India 600036}
\email{sdhomkar@physics.iitm.ac.in}

\author{Basudev Roy}
\affiliation{Department of Physics, IIT Madras, Chennai, 600036, India}
\email{basudev@iitm.ac.in}

\title
  {NaBiF$_4$: Er$^{3+}$, Yb$^{3+}$ upconversion particle as a multi-functional bio-marker}

\abbreviations{IR,NMR,UV}

\begin{document}

\begin{abstract}
Lanthanide-doped upconversion particles (UCPs) have revolutionized optical bioimaging platforms because of their excellent photostability, non-toxicity, and utilization of near-infrared excitation, which facilitates deep tissue penetration with negligible autofluorescence. However, it remains a challenge to achieve high-contrast and sub-diffraction imaging
in noisy biological media, without using a high-power laser. Here, we report various protocols applied to bismuth-doped UCPs address some of these challenges. 
Compared to the photoluminescence (PL) emission of the regular Yttrium doped UCPs, we observe a three-fold increment in the quantum yield of the overall emission of bismuth-UCPs, and a four-fold increment, specifically, in red emission. Leveraging this advantage, we devise a protocol employing two infrared wavelengths, 975 nm and 1064 nm, to selectively control the PL emission. Interestingly, our results reveal two distinct regimes in which PL can be systematically quenched or enhanced, by adjusting the 975 nm laser power. We model the overall dynamics as a simplified stimulated emission depletion process
involving three energy levels. In addition, the particle has a thickness under sub-diffraction, shows optical trapping ability, and potential of surface functionalization to enable specific conjugation with diverse biospecimens.
These studies establish bismuth doped UCPs as an excellent candidate in accomplishing advanced biomarker operating with enhanced signal-to-noise ratio and sub-diffraction imaging capabilities. 
 \end{abstract}

\section{Introduction}

Over the past few years, various agents with enriched strategies have been devised for the development of highly sensitive, robust, and specific biomarkers for disease diagnostics, dynamic cellular monitoring, etc. Conventional biomarking platforms are primarily based on semiconductor quantum dots\cite{Zhang2017BeyondFluorophores,Zhang2023RecentBiomarkers,Martynenko2017ApplicationBiosensing,Chinnathambi2019RecentImaging,Gill2008SemiconductorBioanalysis,Kairdolf2013SemiconductorApplications,Wagner2010UseBiomarkers} or organic fluorophores\cite{Goncalves2009FluorescentProbes,Kim2015DiscoverySeoul-Fluor,Alford2009ToxicityReview}. However, they are constrained by limitations such as photobleaching, fluorescence intermittency, narrow Stokes shifts, and endogenous autofluorescence in biological tissues\cite{Wegner2015QuantumBiosensors,Vu2015QuantumTissue,Abdel-Salam2020SuperiorDots}, restricting the sensitivity and accuracy of the measurements.

Lanthanide-doped upconversion particles (UCPs) have emerged as an attractive alternative to the traditional luminescent biomarkers. These particles absorb in near-infrared wavelength regimes and emit in visible regions through nonlinear anti-Stokes mechanisms, catering to the advantages of deeper penetration ability in biological tissues, minimal autofluorescence\cite{Sordillo2014DeepWindows,Chen2020AdvancedSystems,Cao2020RecentImaging}.
Furthermore, UCP exhibits negligible photobleaching, excellent photostability, and  exceptional steady fluorescence, owing to its specific electronic transitions, which are shielded from the external perturbations.\cite{Haase2011UpconvertingNanoparticles,Chen2014UpconversionTheranostics,Torresan2021CriticalApplications,Wilhelm2015WaterStability}. This facilitates the prolonged and long-range tracking of kinetic processes in biosystems. More recently, the drawback of low quantum yield of UCP has also been addressed through the utilization of different dopants, optimization of dopant concentration, dye-sensitization, and advances in nanostructure engineering, such as core-shell architectures\cite{Boyer2010AbsoluteNanoparticles,Shurukhina2024DependenceIntensity,Wisser2016EnhancingNanoparticles,Kaiser2017Power-dependentSimulations,Wisser2018ImprovingSensitization,Madsen2020ImprovingDesign,Wurth2018QuantumNanoparticles,Wang2011TuningNanoparticles}. 

In addition to their excellent inherent optical properties,
the optical tweezer community has explored various methods for trapping and micro-manipulation of these particles owing to their suitable refractive indices and controllable sizes \cite{Kumar2022EstimationWavelength,Zhang2026UpconversionApplications,Nalupurackal2023SimultaneousMicroparticles,Haro-Gonzalez2013OpticalNanoparticles}. Notably, the same laser of near-infrared wavelength can be used in both trapping and exciting the particles. This dual functionality enables simplified measurement protocols as well as localized probing of heterogeneous and stochastic biological processes which are difficult on an ensemble level. Moreover, as upconversion mechanism requires complex multi-level energy states of the activator and sensitiser, UCPs exhibit saturation and depletion behaviors similar to stimulated emission depletion (STED) processes\cite{Ploschner2020SimultaneousResolution,Krause2019Lanthanide-DopedNanoscopy,Chen2018Multi-photonNanoparticles,Harrington2026UpconvertingLimit,Zhou2026UpconvertingApplications}. This nonlinear dynamics, involving intermediate states, can be controlled employing a second laser through mechanisms such as cross-relaxation, upconversion luminescence depletion or stimulated emission, empowering advanced imaging modalities beyond the diffraction limit.

Although UCPs exhibit low cytotoxicity\cite{Gnach2015UpconvertingToxicity},  precise surface functionalization of the particles is required to be utilized to its full potential as a multifaceted biomarker. Generally, UCP synthesized in a conventional way possesses capping ligands which are hydrophobic in nature, making them unfit in biological mediums. Strategies such as silica coating, amphiphilic polymer encapsulation, ligand exchange have been explored in rendering water solubility, biocompatibility and non-aggregation\cite{Muhr2014UpconversionSurfaces,Bastos2022StabilityCoatings,Himmelsto2019LongTermLigandExchange}. Concurrently, surface engineering facilitates selective recognition of target analytes by attaching specific targeting moieties like peptides, antibodies, aptamers etc. Therefore, functionalized UCP-based biomarkers offer specificity and versatility in complex physiological environments. 

In this work, we perform a comprehensive study of bismuth-doped UCP, NaBiF$_4$: Er$^{3+}$ Yb$^{3+}$ (abbreviated as NaBiF) for biomarker applications. We systematically characterize the structural and optical properties of the particles and, subsequently, 
propose several modalities to harness their potential. 
We also develop a rudimentary model to understand and control the underlying mechanism of intensity dependent PL quenching and enhancement achieved via two-beam pumping.
Furthermore, we carry out preliminary investigations to facilitate the target selectivity and controllability of the marker. Overall, our comprehensive investigation firmly establishes this novel platform as a frontrunner for advance biomarking.

\section{Results and Discussions}

\subsection{Morphological and Optical Characterization}
\begin{figure}[H]
     \centering
     \includegraphics[width=\linewidth]{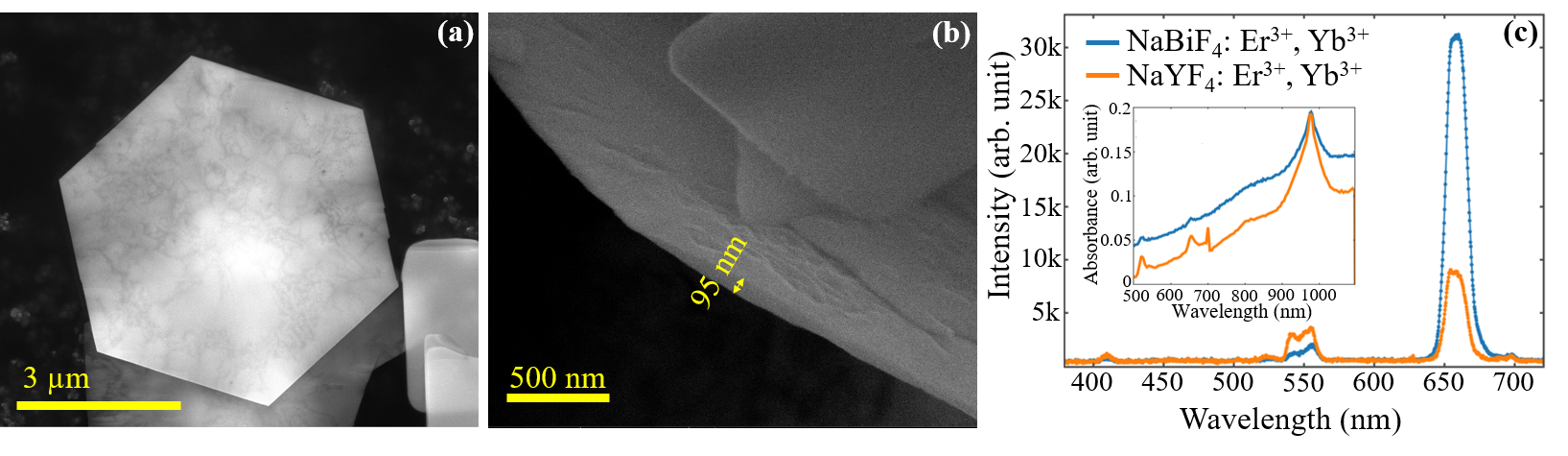}

     \caption{Scanning Electron Microscopy images of the particle in (a) face-on configuration with a diameter of $\sim$ 5.5 $\mu$m, (b) side-on configuration with a thickness of $\sim$ 0.1 $\mu$m. (c) Comparison between photoluminescence spectra of NaBiF and NaYF particles when excited with 975 nm wavelength at 13 mW. Inset: Absorption spectra of the corresponding particles.}
     \label{Figure 1: SEM_PL}
 \end{figure}

We perform a preliminary structural characterization of the NaBiF particles [see Supplementary Information (SI) Section \ref{Synthesis}. for the synthesis details] using field emission scanning electron microscopy (FESEM), as shown in Fig. \ref{Figure 1: SEM_PL}(a) and (b). The particles are nearly transparent and hexagonal in shape, with an average diameter of $\sim$ 6 $\mu$m. 
The geometry is known to be ideal for optical trapping \cite{Nalupurackal2023SimultaneousMicroparticles,Nalupurackal2023ControlledTrap,Ghosh2025DeterminationDirection,Lokesh2022GenerationConfigurations}, however, the thickness ($\sim$ 100 nm) is well within the sub-diffraction regime, promising much richer dynamics to explore and to be leveraged for applications. Additional structural characterization results are shown in SI Fig. \ref{Raman_XRD}.

We further investigate the optical properties of the NaBiF particle and observe that it exhibits stronger emission under 975 nm excitation compared to 1064 nm excitation. Indeed, the UCPs has higher absorption cross section at 975 nm than at 1064 nm, as shown in the Fig. \ref{Figure 1: SEM_PL} (b) inset.  
We compare the brightness of a single NaBiF particle with a single NaYF$_4$: Er$^{3+}$ Yb$^{3+}$ (abbreviated as NaYF) particle under the same conditions and observe $\sim$ 3 times higher quantum yield for NaBiF, in overall emission, than that of the regular NaYF. The red emission of NaBiF particle is $\sim$ 4 times higher than of NaYF particle. Furthermore, the NaBiF particle exhibits significantly higher red fluorescence than green and blue emission.
The observations can be explained from the lattice structure and its impact on the transitions among energy levels. The ionic radius of Bi{$^+$$^3$} ($\sim$ 1.03 \r{A}) and Y{$^+$$^3$} atom ($\sim$ 0.89 \r{A}) is dissimilar, creating 
lattice distortion in the local fields surrounding the host matrix\cite{Verma2020Synchrotron-BasedCrystals,Tian2012MnDelivery}. This significant modification increases the probability of radiative transition, enhancing the luminescence intensity in the red region. Moreover, cross relaxation\cite{Rabouw2018QuenchingNanocrystals,Trave2023LightParticles,Liu2025StrongThermometry,Wang2009UpconversionDependence} between $^4$F$_{7/2}$+ $^4$F$_{9/2}$→$^4$I$_{11/2}$ +$^4$F$_{9/2}$ favors the red emission as opposed to the green or blue. The detailed explanation of the energy transfer mechanism, along with the corresponding energy level diagram, is shown in SI Fig. \ref{Energy levels}.  

\subsection{Surface Functionalization}

Silica (SiO$_\text{2}$) coating is the most common and effective method to crosslink or stabilize UCPs\cite{Kembuan2019CoatingThickness,Iglesias-Mejuto2023SynthesisMethods,Wang2011UpconversionApplications,Kostiv2015SilicacoatedProperties}. Silica forms a rigid, porous, and chemically versatile shell around the UCP through a sol–gel process, creating a stable network that prevents aggregation and enables further functionalization. Details of the surface functionalization method are provided in \ref{Method of Surface functionalization}.

To understand the surface modification of NaBiF particles by silica coating, Fourier
Transform Infrared Spectroscopy (FTIR) characterization of bare NaBiF particle and surface-functionalized NaBiF particles is carried out. FTIR ATR Spectrometer (Thermo Fisher Scientific - Nicolet Summit X) is used
to record the FTIR data at a resolution of 4 cm$^-{^1}$ and we consider the range between 400 and 4000 cm$^-{^1}$in the analysis. The recorded spectra displays characteristic absorption bands associated with the functional groups present before and after functionalization.  Since, NaBiF is an inorganic fluoride material, its spectrum contains only a limited number of infrared-active vibrations, while the lattice vibrations of Na–F and Bi–F are generally located below 500 cm$^-{^1}$.

\begin{figure}[H]
    \centering
    \includegraphics[width=1.0\linewidth]{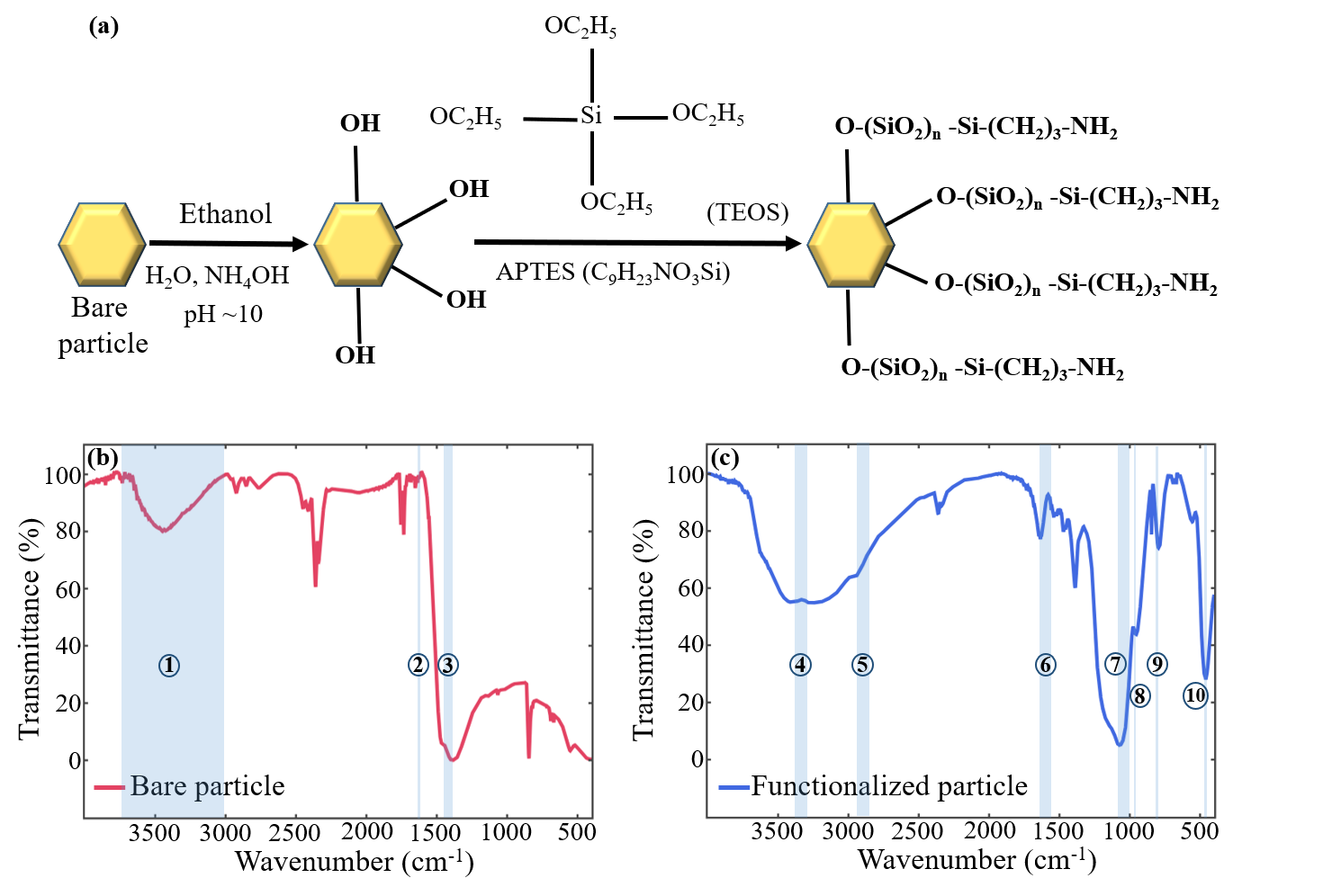}
    \caption{(a) Schematic of the Sequential Surface Modification of a NaBiF particle using TEOS and APTES via hydrolysis and condensation method. (b) FTIR spectra of a bare NaBiF particle, and (c) a surface-functionalized NaBiF particle. }
    \label{Figure 2: FTIR}
\end{figure}

A brief procedure of surface functionalization and the corresponding FTIR data is shown in Fig. \ref{Figure 2: FTIR}, whereas, the origins of the prominent FTIR bands have been compiled in Table \ref{table:FTIR}. The FTIR spectrum of the prepared bare NaBiF nanoparticles shows absorption bands in regions \circled{1}, \circled{2}, and \circled{3}. In the case of a surface-functionalized NaBiF particle, the appearance of prominent bands in the regions \circled{4}, \circled{5}, and \circled{6} indicates the successful deposition of a silica layer around the nanoparticles. A weak absorption near region \circled{7} corresponds to surface silanolgroups, which serve as active sites for further chemical modification. Following treatment with APTES, we observe additional absorption bands around regions \circled{8}, \circled{9}, and \circled{10}. The emergence of these characteristic bands, together with the retention of the silica-related peaks, confirms the successful grafting of amino-functional groups onto the silica-coated nanoparticle surface. The modified surface provides chemically active sites for the covalent attachment of biomolecules, making the nanoparticles suitable for applications such as biosensing, bioimaging, and targeted drug delivery.


\noindent

\begin{table}[htbp]
\centering
\caption{Description of various FTIR bands observed in the data.}
\label{table:FTIR}

\begin{tabularx}{\textwidth}{|c|c|X|}
\hline
No. & FTIR Band (cm$^{-1}$) & \multicolumn{1}{c|}{Origin} \\
\hline
\circled{1} & 3000--3700 & Stretching vibration of hydroxyl groups associated with absorbed water and surface hydroxyl species. \\
\hline
\circled{2} & 1620--1640 & H--O--H bending vibration of physically adsorbed water molecules. \\
\hline
\circled{3} & 1380--1450 & Residual nitrate or carbonate species remaining from the synthesis process or atmospheric CO$_2$ adsorption. \\
\hline
\circled{4} & 1000--1100 & Asymmetric stretching vibration of Si--O--Si bonds. \\
\hline
\circled{5} & 790--810 & Symmetric stretching vibration of the Si--O--Si bond. \\
\hline
\circled{6} & 450--470 & Bending vibration of Si--O--Si. \\
\hline
\circled{7} & 950--970 & Surface silanol (Si--OH) groups. \\
\hline
\circled{8} & 3300--3400 & Stretching vibrations of amino (NH$_2$) groups. \\
\hline
\circled{9} & 1550--1650 & Bending vibrations of amino (NH$_2$) groups. \\
\hline
\circled{10} & 2850--2950 & Aliphatic C--H stretching vibrations of the propyl chains. \\
\hline
\end{tabularx}

\end{table}

\subsection{Optical Trapping ability}

\begin{figure}[H]
     \centering
     \includegraphics[width=\linewidth]{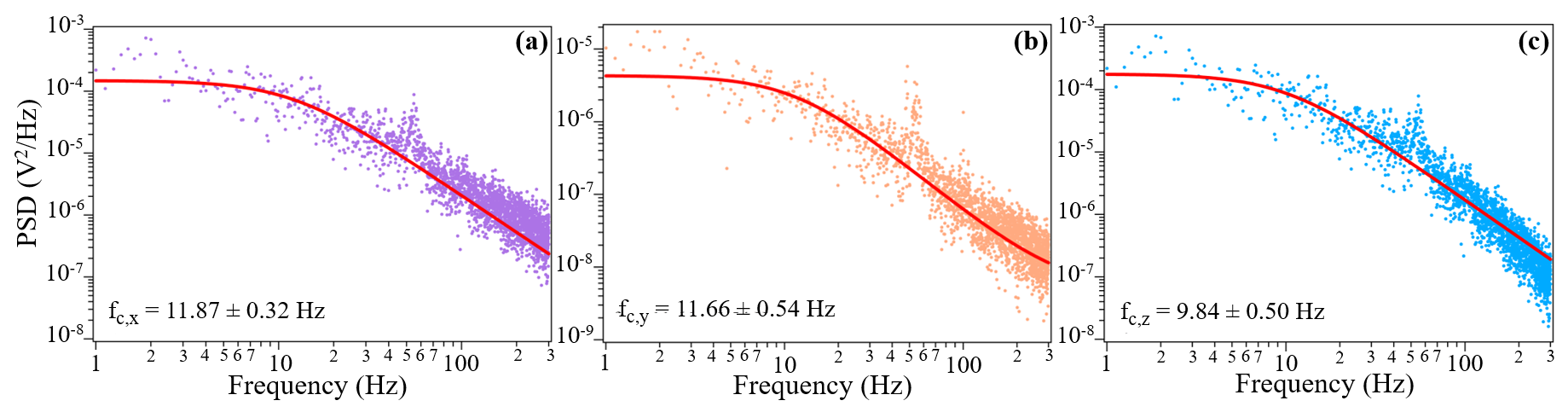}

     \caption{Power spectral density (PSD) of a trapped particle in (a) x-direction, (b) y-direction, and (c) z-direction. The power of the trapping laser is $\sim$ 35 mW in the sample plane. The solid lines represent Lorentzian fits of the form, $PSD_i(f) = \frac{A}{f_{c,i}^2 + f^2}$, where $f_{c,i}$ is the corner frequency along the $i^{\text{th}}$ axis.}
     \label{Figure 3: PSD}
 \end{figure}

The optical trap is built using the optical Tweezer kit (OTKB, Thorlabs), as shown in SI Fig. \ref{schematic}. A single NaBiF particle is trapped with a polarized laser beam of 975 nm wavelength in a water medium, and the forward-scattered light is processed through a quadrant photodiode (QPD) to analyse the trapping efficiencies. The particle is trappable in three dimensions with trap stiffness ($\kappa$) of $\sim$ 0.22 $\mathrm{pN}\,(\mu\mathrm{m}\cdot\mathrm{W})^{-1}$, shown in Fig. \ref{Figure 3: PSD}. In most biological environments with a similar aqueous medium\cite{Roxworthy2014PlasmonicMedia}, it yields a soft but stable optical trapping condition for the controllable transfer of the biomarker.   

\subsection{PL emission control}

We perform pulsed experiments with mutually perpendicular polarizations, using the OTKB setup used for trapping. The particle is kept in a face-on configuration and in a dried condition to observe the effect for a longer period. 

\begin{figure}[H]
    \centering
    \includegraphics[width=0.45\linewidth]{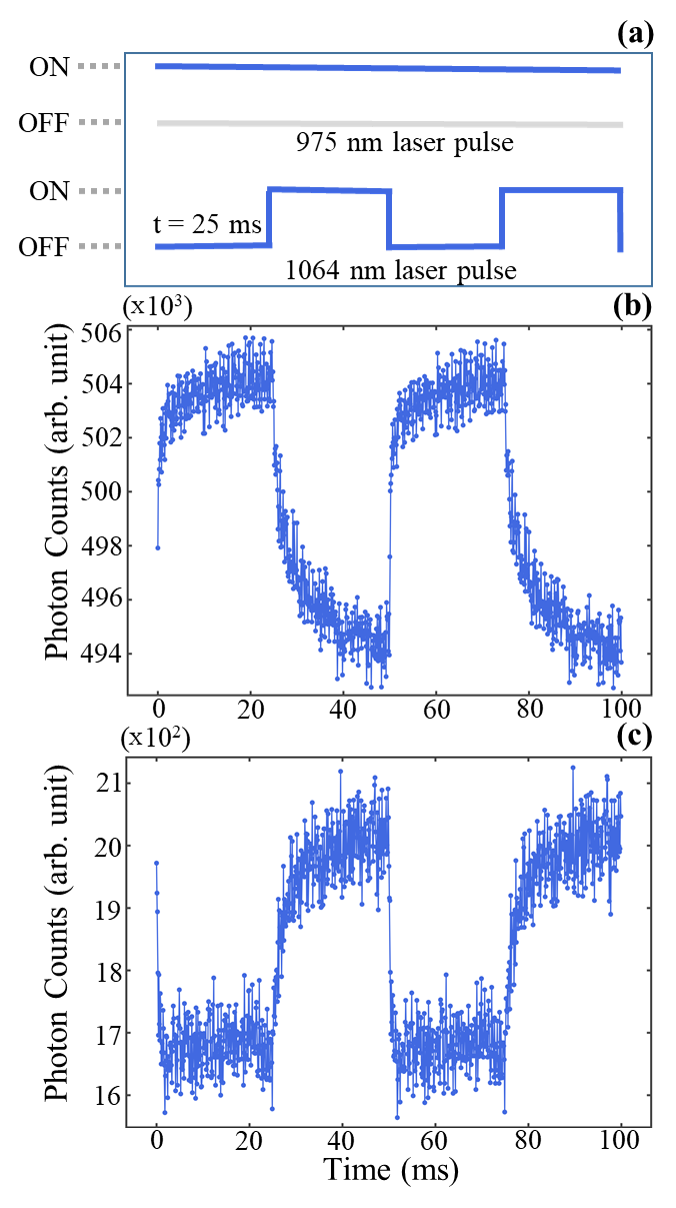}
    \caption{(a) Schematic of the pulse sequence used to probe the quenching dynamics. Experimentally observed effect on PL of a single NaBiF particle upon applying the pulse sequence in two power regimes - (a) the powers of  975 nm and 1064 nm lasers are $\sim$ 19 mW and $\sim$ 77 mW, respectively. (c) the powers of  975 nm and 1064 nm lasers are $\sim$ 0.3 mW and $\sim$ 77 mW, respectively.}
    \label{Figure 4: per_modulation_highP_lowP}
\end{figure}

As the particle shows stronger emission at 975 nm, we keep the laser in CW mode and pulse the 1064 nm laser with a frequency of 20 Hz as shown in Fig. \ref{Figure 4: per_modulation_highP_lowP}(a). The emitted photons are counted by a single-photon detector with a 660 nm bandpass filter to exclusively capture the effect on the prominent red fluorescence. Notably, there are two distinct regimes depending on the power of the 975 nm laser. For a fixed power of 1064 nm, PL quenches at high powers of 975 nm and gets enhanced at comparatively low powers of 975 nm, as shown in Fig. \ref{Figure 4: per_modulation_highP_lowP}(b) and (c), respectively. Although both 975 nm and 1064 nm are known to excite upconversion particles\cite{Chakraborty2025Off-ResonantFe}, the reduction and enhancement of PL upon simultaneous illumination motivated us to build a rudimentary model based on the basic principle of stimulated emission depletion (STED).

\begin{figure}[H]
    \centering
    \includegraphics[width=\linewidth]{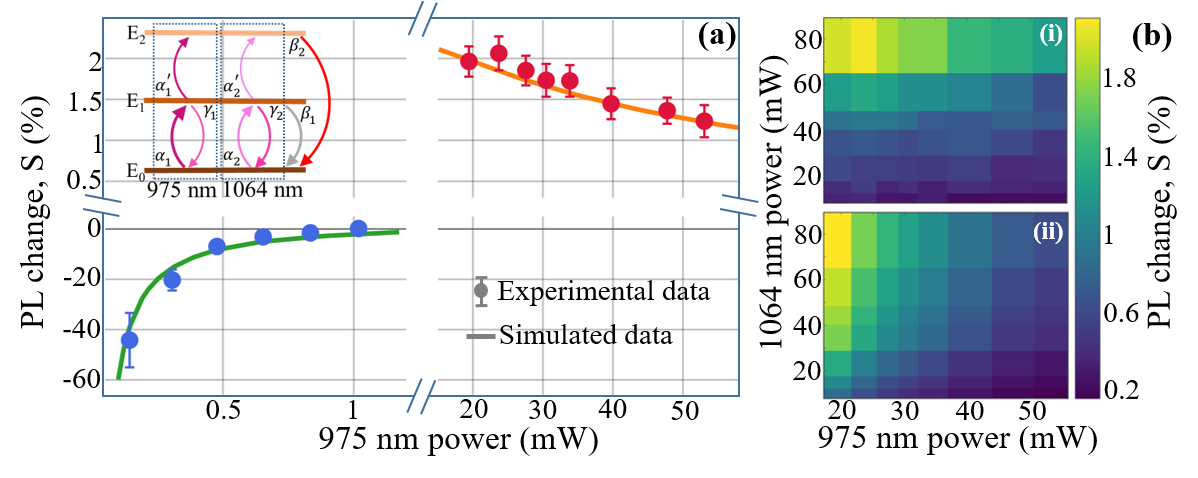      }
    \caption{ (a) The points represent observed PL change when 975 nm power is varied for a fixed power of 1064 nm at 77 mW. Separated by a broken axis, the right side shows the quenching of PL at a high power of 975 nm, and the left side shows the enhancement of PL at a low power of 975 nm. The inset represents the model used to simulate the dynamics. The solid lines are extracted from the simulation. (b) A 2D map of PL change in the high power regime of 975 nm with varied power of 1064 nm - (i) experimental data, (ii) simulated data. }
    \label{Figure 5: 975nm_quenching_statistics}
\end{figure}

We consider an effective three-level closed system to simulate the underlying principle in the upconversion mechanism, as shown in Fig. \ref{Figure 5: 975nm_quenching_statistics} (inset). Three processes govern the overall dynamics: excitation by laser (denoted by $\alpha$ and $\alpha'$ parameters), spontaneous emission (denoted by $\beta$ parameters), and stimulated emission (denoted by $\gamma$ parameters). The timescales of the observed processes are close to the lifetime values of the spontaneous emission (SI Fig. \ref{lifetime975_1064}), leading to the following rate equations.
\begin{align*}
\frac{dN_0(t)}{dt} &=
- \bigl[\alpha_1 I_{975} + \alpha_2 I_{1064}(t)\bigr] N_0(t)
+ \bigl[\beta_1 + \gamma_{1} I_{975} + \gamma_{2} I_{1064}(t)\bigr] N_1(t)
+ \beta_2 N_2(t) ,
\\[6pt]
\frac{dN_1(t)}{dt} &=
\bigl[\alpha_1 I_{975} + \alpha_2 I_{1064}(t)\bigr] N_0(t)
- \bigl[\alpha_1' I_{975} + \alpha_2' I_{1064}(t)
+ \beta_1 + \gamma_{1} I_{975} + \gamma_{2} I_{1064}(t)\bigr] N_1(t) ,
\\[6pt]
\frac{dN_2(t)}{dt} &=
\bigl[\alpha_1' I_{975} + \alpha_2' I_{1064}(t)\bigr] N_1(t)
- \beta_2 N_2(t) .
\end{align*}

Here, $N_0(t)$, $N_1(t)$, and $N_2(t)$ denote the population in the energy levels $E_0$, $E_1$, and $E_2$, respectively at time t, with the constraint of conservation of the total population (normalized) i.e. $N_0(t)$ + $N_1(t)$ + $N_2(t)$ = 1. At time t = 0, the entire population is in the ground state, $N_0(t=0) = 1$, and upon illumination by the laser, it populates the higher energy levels.
Based on the strength of emission and the known energy-level struacture we impose, $(\alpha_1,\alpha_1') >> (\alpha_2,\alpha_2')$ and $\alpha_1<\alpha_1'$, as well as $\alpha_2<\alpha_2'$. As the upconversion mechanism involves metastable intermediate state, it implies the conditions $\beta_2 > \beta_1$. We obtain the value of $\beta_2$ from experimentally measured lifetime. The effective parameters extracted from the optimized model simulations are provided in SI Table \ref{parameter_table}. The parameters re-emphasize the difference in absorption and stimulated emission rates between the two wavelengths. Thus, primarily, the combined effect of these two parameters manifests itself as either quenching or enhancement depending on the illumination intensity regime that we choose to work with. 

Finally, as the experimentally observed upconversion PL originates from the radiative emission of level \(E_2\), the time dependence of $ \beta_2 N_2(t)$ is monitored to probe the effect on emission.
We evaluate the PL change (abbreviated as S) by defining, 

\begin{equation}
S (\%) = \frac{\mu_1 - \mu_2}{\mu_1}\text{ x }100
\end{equation}

$\mu_1$ and $\mu_2$ represent the averaged photon counts when 1064 nm is in off and on conditions, respectively. To exclude initial transient dynamics during pulse transitions, this averaging is restricted to the final three milliseconds of each state.

We vary the laser power to extract the PL behavior as shown in Fig. \ref{Figure 5: 975nm_quenching_statistics}. At a fixed power of 1064 nm, PL quenches slowly in the high-power regime of 975 nm and enhances at a rapidly in the low-power regime. There is a threshold power of 975 nm to achieve the highest quenching percentage, $\sim$ 2\% in the present scenario when the particle is in face-on configuration with the two incident beams in relative perpendicular polarizations.  Additionally, we notice in the heatmaps that quenching percentage can be enhanced with the increment of 1064 nm power,
details are provided in the SI Fig. \ref{quenching_wavelengths_statistics}.

\begin{figure}[H]
    \centering
    \includegraphics[width=\linewidth]{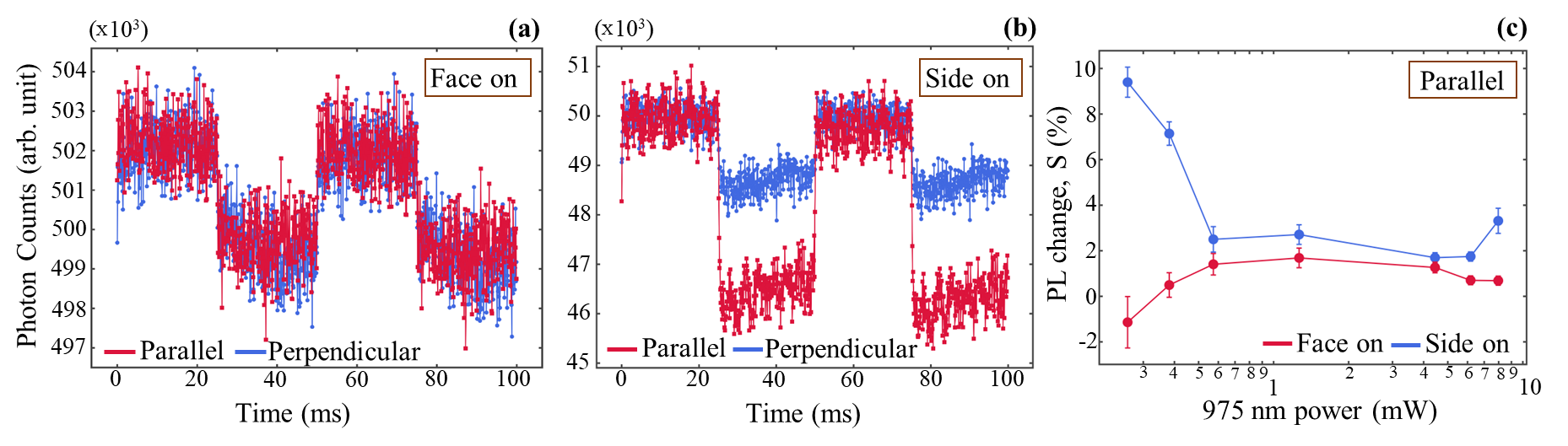}
    \caption{PL change with various configurations of particle orientations and the excitation beam polarization - (a) face on configuration with 975 nm at 24 mW and 1064 nm at 27 mW for both parallel and perpendicular polarizations, (b) side on configuration with 975 nm at 8 mW for both polarizations and 1064 nm at 27 mW and 77 mW for parallel and perpendicular polarizations, respectively (c) Dependence of PL change  in parallel polarization configuration at a fixed power of 1064 nm at 77 mW and a varied power of 975 nm.} 
    \label{Figure 6 : faceOn_sideON_polarization_comparison}
\end{figure}

Interestingly, the quenching percentage can be significantly improved by adjusting the relative polarisations of the two laser beams with the physical configuration of the particle, as shown in Fig. \ref{Figure 6 : faceOn_sideON_polarization_comparison}. The particle can be considered as a neatly arranged electrical dipoles when seen from the side and those placed in random orientations when seen from the top\cite{Rodriguez-Sevilla2016DeterminingSpectroscopy}. As a result, relative parallel and perpendicular configurations of the incident beams does not affect when particle is faced on, whereas show considerable effect in side-on configuration. We notice, with the parallel polarization, the magnitude as well as nature of PL change can be substantially regulated, achieving a high $\sim$ 10\% quenching at relatively low powers of 975 nm for the side-on configuration. This is particularly advantageous for achieving optical trapping at low laser powers, a key requirement in biological environments where minimizing heating \cite{Chakraborty2023FacetsParticles} and reducing water absorption effects \cite{Semak2019MeasurementWavelength,Chakraborty2025Off-ResonantFe} is highly desirable.

\noindent\textbf{Potentialities: Lock-in measurement, sub-diffraction imaging, and temperature sensing}

Our experiments suggest that there are several avenues for the future exploration of Bi$^{\text{3+}}$ doped UCP. We fit the quenching and enhancing dynamics with double exponential and find that one of the lifetimes is in the order of 5 ms (see SI Fig. \ref{lifetime975_1064} and Table \ref{double_exponential_statistics}). This prolonged duration is characteristic of the intrinsic excited-state lifetime of the $\text{Yb}^{3+}$ sensitizer (specifically the $^2\text{F}_{5/2}$ state)\cite{Vonk2024RiseNanocrystals}, which acts as a crucial intermediate energy reservoir. The timescales involved in the PL change process are independent of laser powers as well as the temperature (SI Fig. \ref{dbEXP_statistics}). Additionally, the fluorescence lifetime of the red emission is nearly 0.2 ms. Employing lock-in detection techniques\cite{Zheng2016High-ContrastApproach,Bagheri2025LanthanideImaging,Labrador-Paez2023Frequency-DomainKinetics}, this long-lived upconversion signal can be selectively extracted while suppressing short-lived background of highly autofluorescent or scattering bio-environments. Importantly, both the excitation and emission wavelengths are situated within the near-infrared biological optical windows\cite{Chen2014UpconversionTheranostics}. This strategic spectral positioning minimizes photon attenuation from tissue scattering and absorption, ultimately amplifying the signal-to-noise ratio (SNR) by several orders of magnitude for deep-tissue applications.

Because of the inherent multi-photon processes, UCPs breach the optical diffraction limit \cite{Chen2022ExploitingImaging} as illustrated by the simulations in SI Fig. \ref{Theoretical_resolution}. Additionally, introducing Bi$^{\text{3+}}$ doping into the UCP architecture provides the critical brightness required for sub-diffraction resolution. Moreover, our 2-beam experiments suggest that there is a possibility of reducing the power of STED beam significantly, which would be highly desirable to avoid specimen damage. 

Lastly, UCPs have also been employed as local temperature sensors \cite{Jha2024-NaBiF4:Yb3+Er3+:NovelThermometry,Antoniak2020CombinedNanoparticles}. Our temperature dependent measurements shown in SI Fig. \ref{temperature_sensing} illustrate a novel approach to thermometry that utilizes PL change observed in the 2-beam experiments as a reliable proxy for temperature sensing. More work is required to study and optimize the accuracy of the sensor.

\section{Summary}
In summary, we present several modalities on harnessing bismuth-doped upconversion particles as proficient sub-diffraction biomarkers. The particle has a thickness in the nanometer regime and a shape suitable for optical trapping. We show that the particles can be functionalized for targeted applications. Leveraging its bright emission, we select a specific red window and show that PL can be modulated controllably by applying two infrared wavelengths. The PL can be quenched and enhanced efficiently by selecting the relative polarizations of the incident beams, their powers and the orientation of the particle. Furthermore, we present a simplified model on the governing dynamics as STED process. Collectively, these approaches position bismuth-doped UCPs as a formidable, multi-functional tool for probing biochemical processes.

\begin{acknowledgement}
We thank the Indian Institute of Technology, Madras, India for their seed and initiation grants to BR. This work was also supported in parts by the DBT/Wellcome Trust India Alliance Fellowship IA/I/20/1/504900 awarded to BR. S.D. thanks the Indian Institute of Technology, Madras, India, and the Science and Engineering Research Board (SERB Grant No. SRG/2023/000322), India, for start-up funding. S.D. and A.G. acknowledge the use of the computational facilities supported by a grant from the Mphasis F1 Foundation given to the Center for Quantum Information, Communication, and Computing (CQuICC).
\end{acknowledgement}

\noindent\textbf{\large{Disclosures}  }

\noindent There are no conflicts of interest to declare. 

\noindent\textbf{\large{Data and Code Availability}  }

\noindent Data and the code underlying the results presented in this paper may be obtained from the authors upon reasonable request.

\newpage
\bibliography{references.bib}

\clearpage


\renewcommand{\thesubsection}{S\arabic{section}.\arabic{subsection}}

\setcounter{figure}{0}
\renewcommand{\thefigure}{S\arabic{figure}}

\setcounter{table}{0}
\renewcommand{\thetable}{S\arabic{table}}

\setcounter{equation}{0}
\renewcommand{\theequation}{S\arabic{equation}}

\section*{Supplementary Information (SI)}

\makeatletter
\subsection*{SI.1. Synthesis of NaBiF$_4$: Er$^{3+}$, Yb$^{3+}$ Particles}
\phantomsection
\def\@currentlabel{SI.1}\label{Synthesis}
\makeatother

Synthesis of NaBiF particles is done in hydrothermal method. Bismuth nitrate Bi(NO3), Yttrium(III) nitrate hexahydrate  [Y(NO$_3$)$_3$.6H$_2$O], Ytterbium(III) nitrate hexahydrate 
[Yb(NO$_3$)$_3$.6H$_2$O], Erbium(III) nitrate hexahydrate [Er(NO$_3$)$_3$.6H$_2$O], and sodium citrate dihydrate (Na$_3$C$_6$H$_5$O$_7$), Sodium Floride (NaF) are procured from Alfa Aesar. Doubly distilled (DI) water( Millipore DQ 3, MerckSystems) and ethanol (C$_2$H$_5$OH) (Hayman, 99.9\%) are used along the synthesis process. 

1.23 g of sodium citrate and 1.592 g of Bi(NO$_3$) 0.0328 M are added to 20 ml of water in a clean beaker, and then the solution is vigorously stirred until it becomes transparent. Then, 0.04 g of Er(NO$_3$)$_3$·6H$_2$O, and 0.38 g of Yb(NO$_3$)$_3$·6H$_2$O are added to the above solution. Subsequently, 17 ml of ethanol is added to the above mixture and stirred for 5 min. After stirring, 0.121g of CTAB is added to the above solution and further stirred for 15 minutes. In a separate beaker, 1.411 g of NaF and 67 ml of DI water are added and stirred for a few minutes until a clear solution is obtained. The NaF solution is slowly mixed with the rare earth solution to obtain a transparent solution. The whole mixture is stirred for 2 hours at a mild speed of 700 rpm. Then, the solution is transferred to a Teflon tube, autoclaved, and kept for 12 hours at 200 degrees Celsius.

\makeatletter
\subsection*{SI.2. Method of Surface functionalization}
\phantomsection
\def\@currentlabel{SI.2}\label{Method of Surface functionalization}
\makeatother

50 mg of NaBiF$_4$ and 20 mL of ethanol are taken in a 50 mL beaker and is stirred for 5–10 min to obtain a uniform dispersion. Subsequently, 1 mL of deionized (DI) water and 0.5 mL of 25\% NH$_4$OH are added to the mixture. Thereafter, 100 $\mu$L of TEOS is added under vigorous stirring. For amine functionalization, 30 $\mu$L of APTES is added simultaneously with TEOS. The resulting solution is then stirred at room temperature for 18 h (a longer reaction time is required to obtain a thicker shell). The functionalized particles are then washed with ethanol to terminate the reaction, followed by repeated centrifugation and washing with deionized water to remove residual ammonia and unreacted silane species.

\makeatletter
\subsection*{SI.3. Additional Structural Characterizations}
\phantomsection
\def\@currentlabel{SI.3}\label{Additional Structural Characterizations}
\makeatother

\begin{figure}[H]
     \centering
     \includegraphics[width=\linewidth]{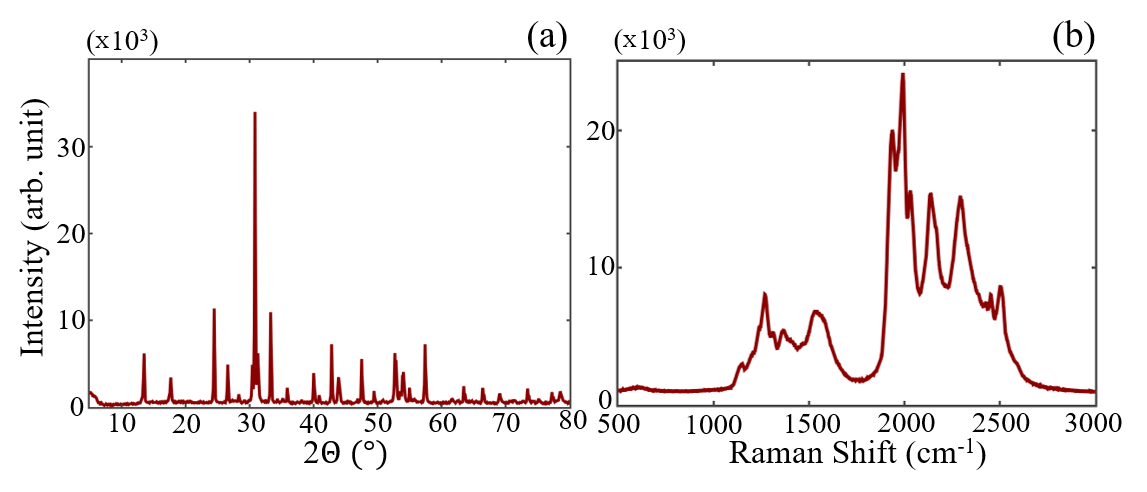}
     \caption{(a) XRD, and (b) Raman spectra of NaBiF$_4$: Er$^{3+}$, Yb$^{3+}$ particles.}
     \label{Raman_XRD}
 \end{figure}

 The crystal structure of the prepared sample is analyzed using XRD characterisation, shown in Fig. \ref{Raman_XRD}(a). The diffraction peaks were well matched with standard hexagonal crystal structure of NaBiF (JCPDF card no. 41-0796). The peak position appeared at 2 theta angle 17$^0$, 30.7$^0$,33.3$^0$, 42.8$^0$, 49.5$^0$, 52.6$^0$, 57.4$^0$, 69.1$^0$ corresponding to hkl value of (100), (110), (200), (201), (002), (211), (112), (311) value. The sharp intensity peaks resemble the good crystallinity of the materials. High intensity peak appeared at 30.7 degree shows dominant hexagonal phase of the prepared sample.  Introducing rare earth dopant (Yb,Er) in the crystal matrix did not the change the compound and phase of the materials. This shows the Yb and Er dopants successfully replaced the Bi sites. The remaining peak observed in addition to NaBiF phase is corresponds to hybrid structure of BiF3 and BiOF impurity phase.

Raman spectrum used to depicts the vibrational phonon modes of NaBiF$_4$: Yb$^{3+}$, Er$^{3+}$ was shown in fig \ref{Raman_XRD}(b). The intensity peak observed at  $1274$ cm$^{-1}$, $1365$cm$^{-1}$ and $1554$cm$^{-1}$ belongs to host lattice NaBiF$_4$ phonon vibrations. Peak intensity observed between $1939$cm$^{-1}$ to $2054$cm$^{-1}$ is strain induced by doping of Yb$^{3+}$, Er$^{3+}$ in Bi sites and also carboxyl group present in the sample.

\makeatletter
\subsection*{SI.4. Upconversion process}
\phantomsection
\def\@currentlabel{SI.4}\label{upconversion process}
\makeatother

\begin{figure}[H]
    \centering
    \includegraphics[width=0.5\linewidth]{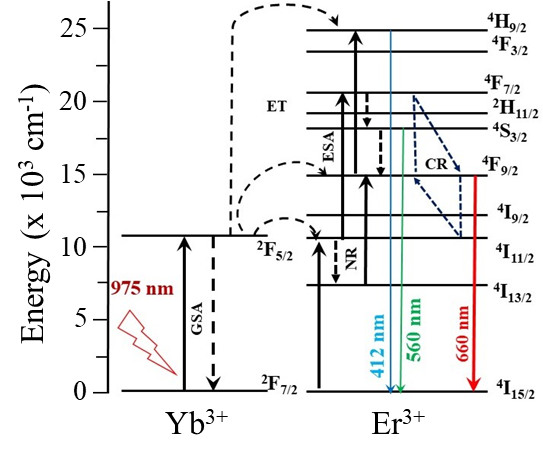}
    \caption{Energy level diagram of NaBiF$_4$:Yb{$^3$$^+$}, Er{$^3$$^+$} upconversion particle and possible transitions pathways.}
    \label{Energy levels}
\end{figure}

\begin{figure}[H]
\centering\includegraphics[width =\linewidth]{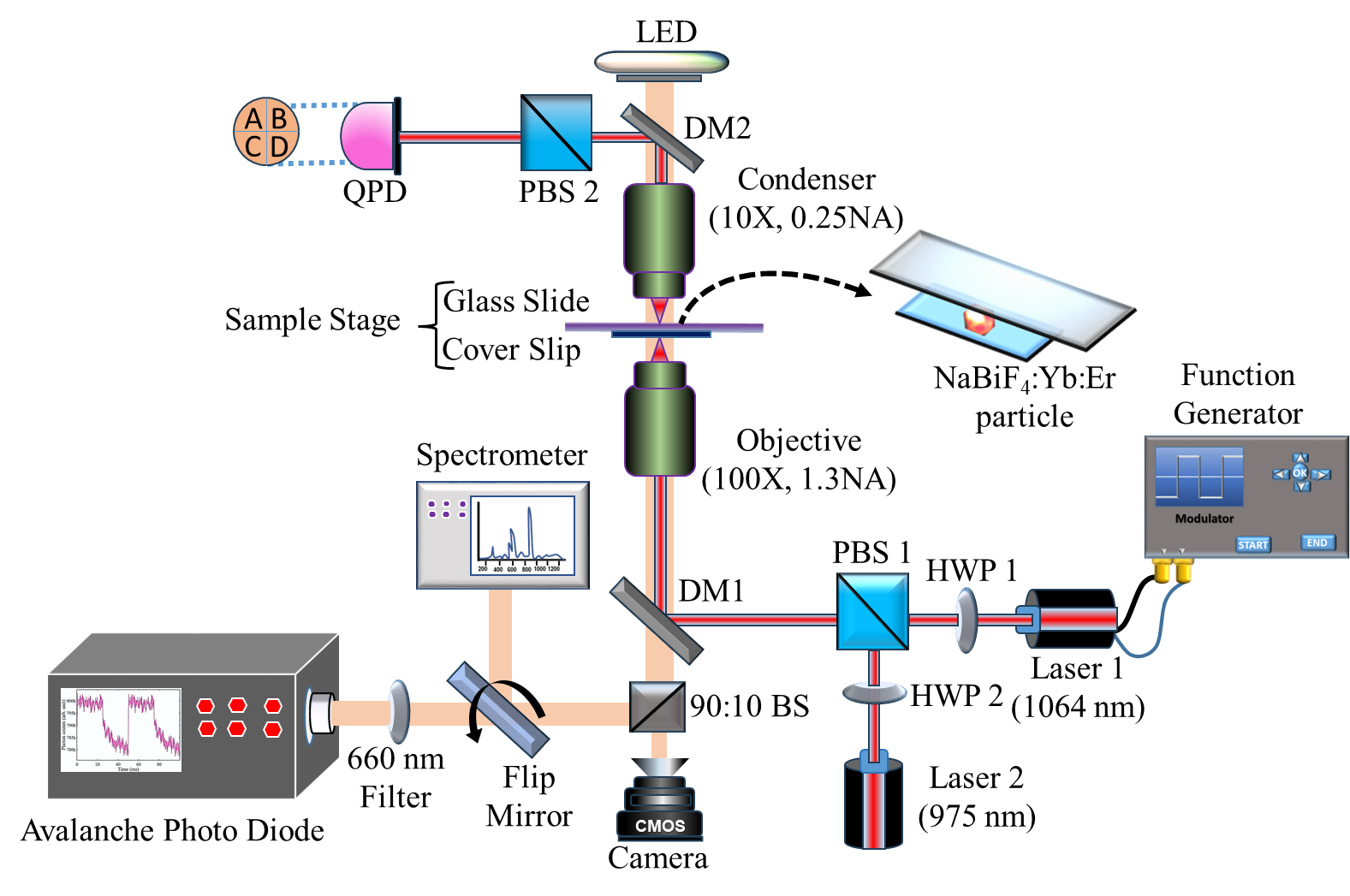}
\caption{Schematic diagram of the experimental setup, representing excitation with two infrared wavelengths in specific polarization and collection at the backscattering direction by spectrum analyzer and avalanche photodiode.}
\label{schematic}
\end{figure}

\begin{figure}[H]
    \centering
    \includegraphics[width=\linewidth]{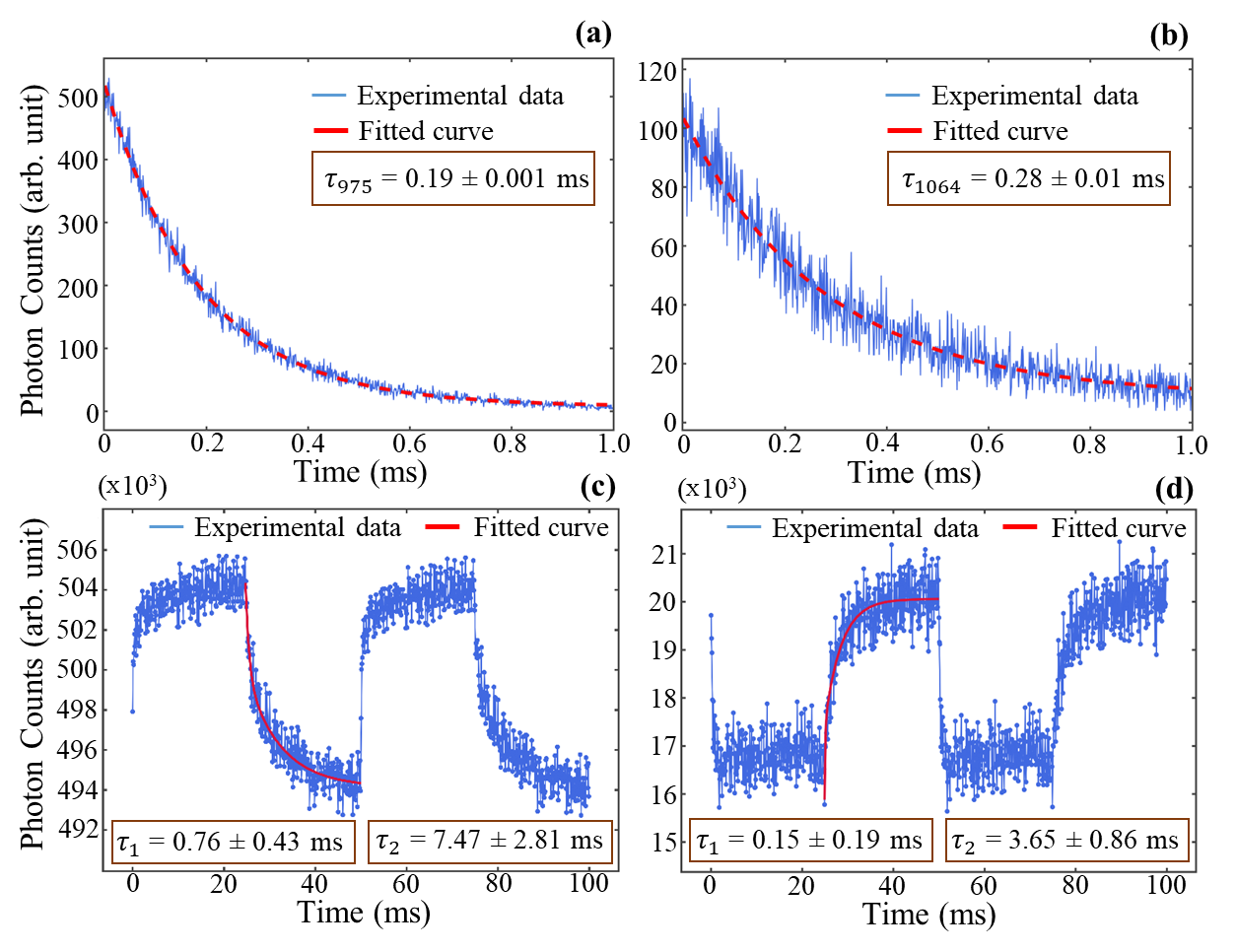}
    \caption{Figures showing lifetimes in the presence of (a) 975 nm wavelength at 1.3 mW,  (b) 1064 nm wavelength at 27.6 mW. In (c) and (d), we fit the data with a bi-exponential of the functional form, $f(t) = A_1 e^{-t/\tau_1} + A_2 e^{-t/\tau_2} + C$ to extract the timescales involved in the respective processes.}
    \label{lifetime975_1064}
\end{figure}

\begin{figure}[H]
    \centering
    \includegraphics[width=\linewidth]{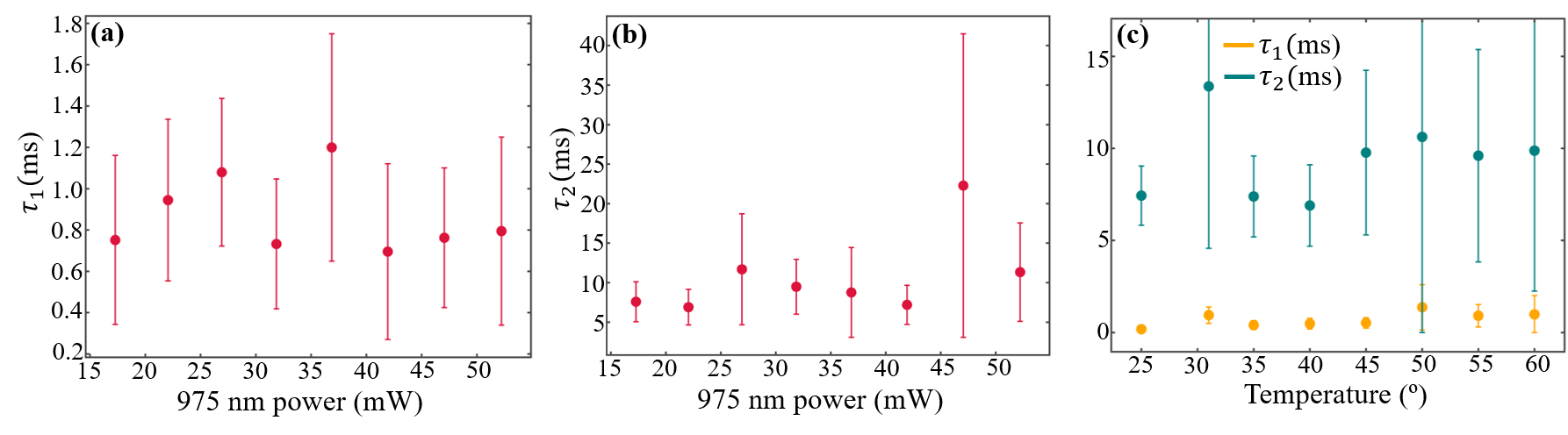}
    \caption{The statistical analysis of how the two timescales depend on the laser powers and temperatures - (a) $\tau_1$ and (b) $\tau_2$. Here, 1064 nm is kept fixed at 77 mW while 975 nm laser powers have been varied. (c) 1064 nm and 975 nm lasers are kept fixed at 77 mW, and 53 mW, respectively.}
    \label{dbEXP_statistics}
\end{figure}

\begin{figure}[H]
    \centering
    \includegraphics[width=\linewidth]{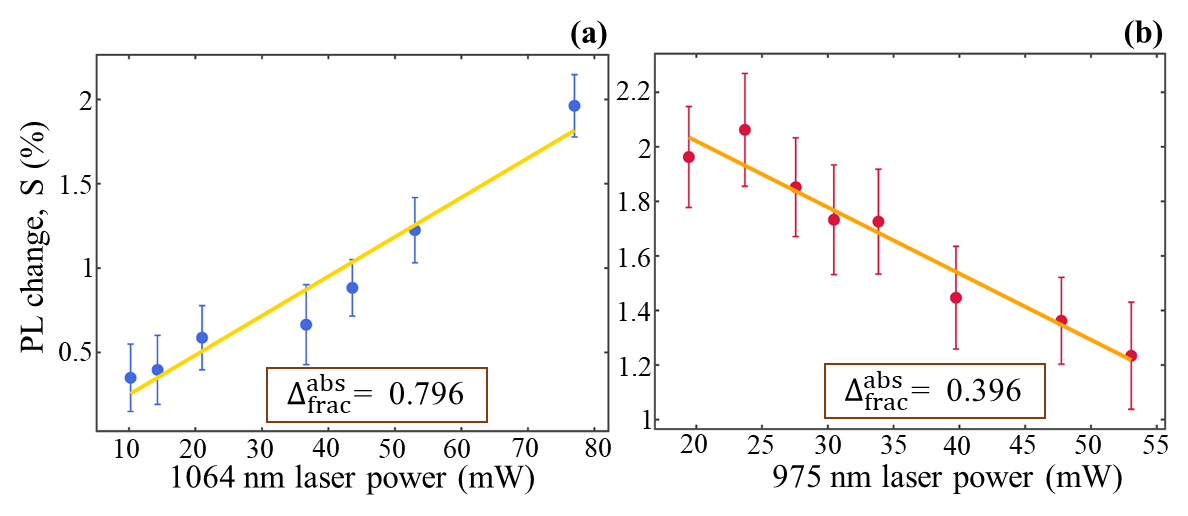}
    \caption{The effect of incident laser power on the PL change. To measure the intrinsic dependency, we calculate $
\Delta_\mathrm{frac}^{\text{abs}} = \frac{|dS/dP| \cdot \Delta P}{S_\mathrm{max}}
$, S$_\text{max}$ represents the PL change at maximum power.}
    \label{quenching_wavelengths_statistics}
\end{figure}

\begin{figure}[H]
    \centering
    \includegraphics[width=\linewidth]{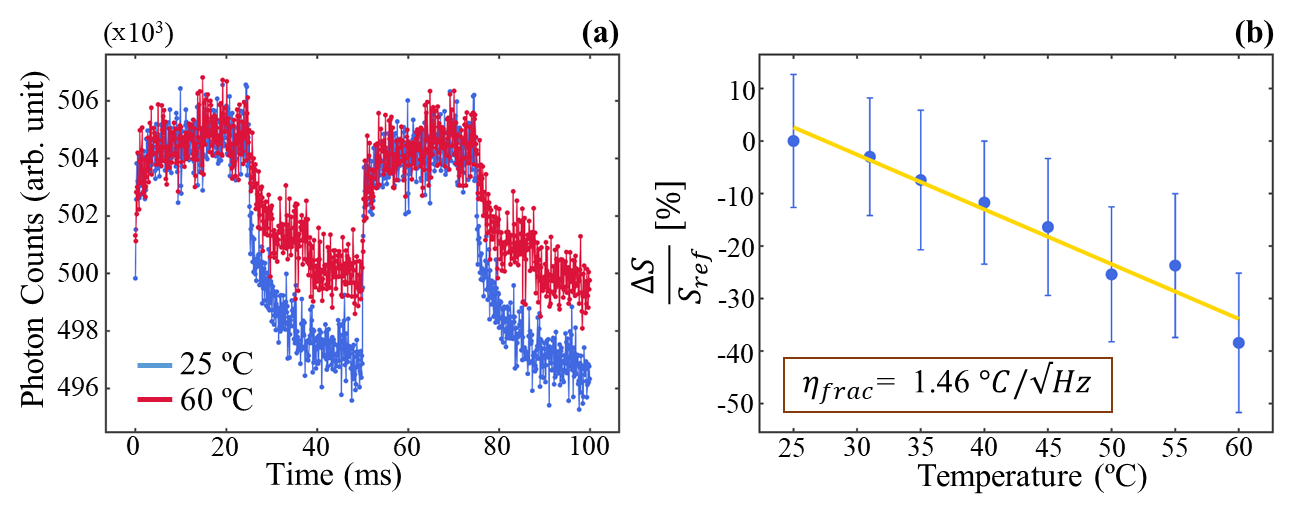}
    \caption{(a) The temperature dependence of the PL change. (b) Calculated temperature sensitivity. Here, S$_{\text{ref}}$ is taken at 25$^\circ$C and  $\eta_f = \frac{\sigma_f}{|m_f|} \sqrt{\tau} \quad (\mathrm{^\circ C/\sqrt{Hz}})$, where $\tau$ is the integration time, $m_f$ is the slope and $\sigma_f$ is standard deviation of the residual from the fitting.}
    \label{temperature_sensing}
\end{figure}

\begin{figure}[H]
    \centering
    \includegraphics[width=\linewidth]{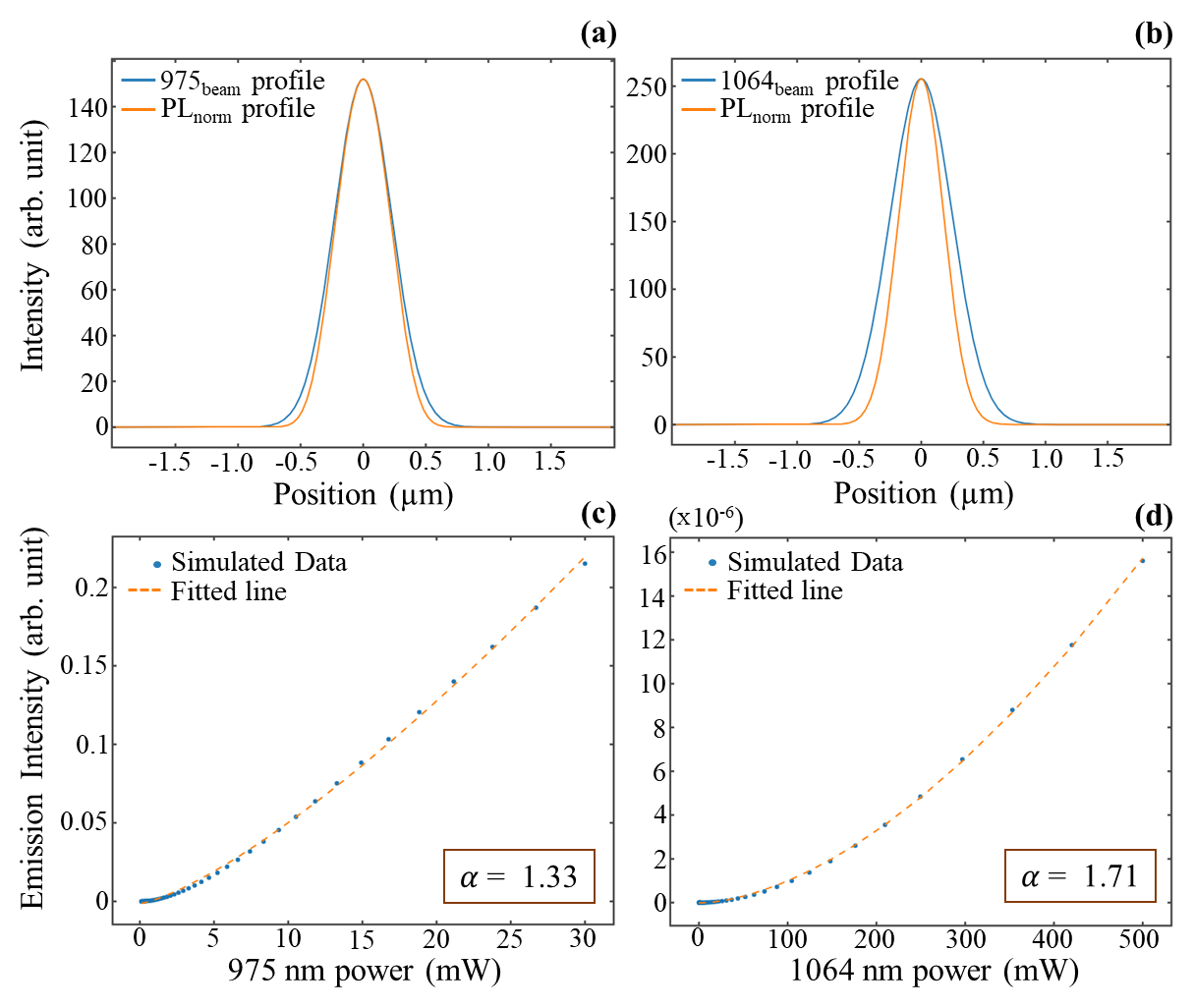}
    \caption{ Incident beam profile and the normalized PL profile in upconversion emission, obtained from simulation - (a) when excited with a Gaussian 975 nm laser at 50 mW power, the FWHM of the beam of 0.54 $\mu$m shrinks to FWHM$_{\text{PL emission}}$ of 0.49 $\mu$m, and (b) for excitation with a Gaussian 1064 nm laser with 100 mW power, we obtain FWHM$_{\text{1064 beam}}$ = 0.58 $\mu$m and FWHM$_{\text{PL emission}}$ = 0.42 $\mu$m, a much higher shrinking. (c) and (d) The quadratic power law behavior of the PL emission. The emission is well within the saturation region for 1064 nm excitation and approaching saturation for 975 nm.}
    \label{Theoretical_resolution}
\end{figure}

The mechanism involved in the NaBiF$_4$:Yb{$^3$$^+$}, Er{$^3$$^+$} upconversion luminescence process could be illustrated from the energy level diagram shown in figure \ref{Energy levels}. Upon excitation with NIR laser (975nm), the photon gets absorbed (sensitizer) Yb{$^3$$^+$} ($^2$F$_{7/2}$) and excited to ($^2$F$_{5/2}$) state by ground state absorption process (GSA). The excited energy gets transferred from Yb{$^3$$^+$} ($^2$F$_{5/2}$) to Er{$^3$$^+$} (activator) $^4$H$_{9/2}$, $^4$F$_{9/2}$ and $^4$I$_{11/2}$ through energy transfer upconversion process (ETU).In addition to this process, the ground state Er{$^3$$^+$}ion also gets excited from $^4$I$_{15/2}$ → $^4$I$_{11/2}$ under 975nm excitation by GSA process. The more population at $^4$I$_{11/2}$ state further get excited to $^4$F$_{7/2}$ by absorbing another photon by two photon absorption process and partially de-excited to $^4$I$_{13/2}$ level by nonradiative relaxation process (NR). The state at $^4$I$_{13/2}$ of Er{$^3$$^+$} further excited to $^4$F$_{9/2}$ by absorbing neighbouring Er ions resulting radiative transition of red emission at 660nm ($^4$F$_{9/2}$→ $^4$I$_{15/2}$). From the energy state at $^4$F$_{7/2}$ of Er decays non radiatively to $^4$S$_{3/2}$ through  $^2$H$_{11/2}$ resulting green emission at 560nm by $^4$S$_{3/2}$→ $^4$I$_{15/2}$ transition. Furthermore, blue emission at 410nm was occurred by the transition of $^2$H$_{9/2}$→$^4$I$_{15/2}$ level. 

Compared to green and blue emission, the red emission is getting more dominant in NaBiF$_4$ phosphor material. The probability of getting high red luminescence intensity is due to two ways of transition occur between Yb{$^3$$^+$} → Er{$^3$$^+$} ions. First way: $^4$I$_{13/2}$+$^4$F$_{9/2}$ →$^4$I$_{15/2}$ and also direct transition of Yb{$^3$$^+$} $^2$F$_{5/2}$ → Er{$^3$$^+$} $^4$F$_{9/2}$ levels. Second way: $^4$I$_{11/2}$ + $^4$F$_{7/2}$→$^4$F$_{9/2}$ + $^4$I$_{15/2}$ through multi phonon relaxation process. The cross relaxation between $^4$F$_{7/2}$+ $^4$F$_{9/2}$→$^4$I$_{11/2}$ +$^4$F$_{9/2}$ states make the transition more at red emission rather than green and blue.

\makeatletter
\subsection*{SI.5. Experimental Method in Optical Tweezer}
\phantomsection
\def\@currentlabel{SI.5}\label{Experimental Method in Optical Tweezer}
\makeatother

The lasers used for excitation are polarized by half-wave plates and passed through a polarizing beam splitter (PBS). These polarized beams are made to fall onto a dichroic mirror and then towards an objective lens (1.3 NA, $100 \times$, Olympus, oil immersion), which tightly focuses the beam into the sample stage. The samples are kept between a glass coverslip of $\sim$170 $\mu$m and a glass slide of $\sim$ 1 mm, with the coverslip at the bottom.

To excite the upconversion particle, two lasers of infrared wavelengths (IR) are used in orthogonal polarization: one of 975 nm (Butterfly laser, Thorlabs) and the other 1064 nm (diode laser, Lasever). The particle emits in the visible range, which is passed through a dichroic mirror to a 90:10 beam splitter. A CMOS camera in the back focal plane is used for imaging, capturing 10 percent of the light, while the rest is directed to collect the spectrum and to count photons. A visible LED source is used to illuminate the sample chamber from the top. The visible emission is guided through a flip mirror to pass to a spectrum analyzer (Research India, RI Series Spectrometer,  Model No. RIFS-VSS-TEC) or to an avalanche photo diode (APD, Excelitas, Model No. SPCM-AQRH-13-FC). As the emission spectrum consists of wavelengths in red, green, and blue, we keep a 661 nm band-pass filter before APD to filter out the prominent red fluorescence. An infrared wavelength blocker is also placed in front of the APD to block any unintended laser light from the collection path, capturing only the emission. We connect time-correlated single photon counter (Time Tagger 20, Swabian Instruments) to the APD and record the number of photons for analysis. 

For performing measurements on the quenching of fluorescence, we pulse the laser of 1064 nm wavelength with a function generator (GW Instek, AFG-2225) and maintain the laser of 975 nm wavelength in continuous wave (CW) mode. The Histogram measurement class in the Time Tagger Module is used to obtain data values with synchronized triggering, utilizing the same pulse as that for the 1064 nm wavelength.

\makeatletter
\subsection*{SI.6. Parameters used in the modelling}
\phantomsection
\def\@currentlabel{SI.6}\label{Parameters used in the modelling}
\makeatother

\begin{table}[H]
    \centering
    \begin{tabular}{|c|c|} 
     \hline
     Parameters & Effective Values \\ 
     \hline
     $\beta_{1}$ & $\frac{1}{0.5}$ $m s^{-1}$\\
     $\beta_{2}$ & $\frac{1}{0.2}$ $m s^{-1}$\\
     $\alpha_{1}$ & $1.4 \times 10^{-2}$ $m W^{-1}m s^{-1}$\\
     $\alpha_{1}'$ & $1.6 \times 10^{-2}$ $m W^{-1}m s^{-1}$\\
     $\gamma_{1}$ & $3.9 \times 10^{-2}$ $m W^{-1}m s^{-1}$\\
     $\alpha_{2}$ & $4.5 \times 10^{-6}$ $m W^{-1}m s^{-1}$\\
     $\alpha_{2}'$ & $7.0 \times 10^{-6}$ $m W^{-1}m s^{-1}$\\
     $\gamma_{2}$ & $1.0 \times 10^{-3}$ $m W^{-1}m s^{-1}$\\
     \hline
    \end{tabular}
    \caption{The table shows effective values for the parameters used in the simulated model to match the experimental datas approximately.}
    \label{parameter_table}

\end{table}

\makeatletter
\subsection*{SI.7. Unaltered lifetimes of two processes}
\phantomsection
\def\@currentlabel{SI.7}\label{Unaltered lifetimes of two processes}
\makeatother

\begin{table}[H]
\centering
\label{tab:avg_lifetimes}
\begin{tabular}{lcc}
\hline
Varied parameter & averaged $\tau_1$ (ms) & averaged $\tau_2$ (ms) \\
\hline
975 nm power & 0.87 $\pm$ 0.14 & 10.65 $\pm$ 2.85 \\
1064 nm power & 0.71 $\pm$ 0.24 & 11.54 $\pm$ 25.57 \\
Temperature & 0.72 $\pm$ 0.23 & 9.37 $\pm$ 2.25 \\
\hline
\end{tabular}
\caption{Average lifetimes and their corresponding standard errors.}
\label{double_exponential_statistics}
\end{table}

\makeatletter
\subsection*{SI.8. Enhancement of quenching}
\phantomsection
\def\@currentlabel{SI.8}\label{Enhancement of quenching}
\makeatother

\makeatletter
\subsection*{SI.9. Temperature Sensing}
\phantomsection
\def\@currentlabel{SI.9}\label{Temperature Sensing}
\makeatother

\makeatletter
\subsection*{SI.10. Obtainable resolution}
\phantomsection
\def\@currentlabel{SI.10}\label{Obtainable resolution}
\makeatother

\end{document}